# Quantitative CT texture-based method to predict diagnosis and prognosis of fibrosing interstitial lung disease patterns

Babak Haghighi[1], Warren B. Gefter[1], Lauren Pantalone[1], Despina Kontos[1], Eduardo J. Mortani Barbosa Jr.[1*]

1: Perelman School of Medicine, University of Pennsylvania, Philadelphia, PA, USA.

[*]**Corresponding author:**
Dr. Eduardo J. Mortani Barbosa, Jr., Address: Division of Cardiothoracic Imaging, Department of Radiology, Perelman Scholl of Medicine, University of Pennsylvania, 3400 Spruce Street, Ground Floor Founders Bldg. Philadelphia, PA, USA 19104
Email: Eduardo.Barbosa@pennmedicine.upenn.edu

**Abstract**

**Purpose:** To utilize high-resolution quantitative CT (QCT) imaging features for prediction of diagnosis and prognosis in fibrosing interstitial lung diseases (ILD).

**Approach:** 40 ILD patients (20 usual interstitial pneumonia (UIP), 20 non-UIP pattern ILD) were classified by expert consensus of 2 radiologists and followed for 7 years. Clinical variables were recorded. Following segmentation of the lung field, a total of 26 texture features were extracted using a lattice-based approach (TM model). The TM model was compared with previously histogram-based model (HM) for their abilities to classify UIP vs non-UIP. For prognostic assessment, survival analysis was performed comparing the expert diagnostic labels versus TM metrics.

**Results:** In the classification analysis, the TM model outperformed the HM method with AUC of 0.70. While survival curves of UIP vs non-UIP expert labels in Cox regression analysis were not statistically different, TM QCT features allowed statistically significant partition of the cohort.

**Conclusions:** TM model outperformed HM model in distinguishing UIP from non-UIP patterns. Most importantly, TM allows for partitioning of the cohort into distinct survival groups, whereas expert UIP vs non-UIP labeling does not. QCT TM models may improve diagnosis of ILD and offer more accurate prognostication, better guiding patient management.

**Abbreviations List:**

UIP: usual interstitial pneumonia

NSIP: nonspecific interstitial pneumonia

CHP: chronic hypersensitivity pneumonitis

HM: histogram-based method

TM: texture-based method

QCT: quantitative computed tomography

HRCT: high resolution computed tomography

ILD: fibrosing interstitial lung diseases

CVs: clinical variables

ROI: region of interest

## Main Manuscript

## Introduction

Accurate classification and quantification of severity of chronic fibrosing interstitial lung diseases (ILD) is crucial for prognostication, management and assessment of treatment response [1, 2, 3, 4, 5]. ILDs are a heterogeneous group of diseases characterized by interstitial inflammation and variable degrees of fibrosis, with widely different morbidity and mortality, often leading to reduced lung volume, decreased lung compliance, restrictive physiology, and potentially, respiratory failure. Progressive fibrotic ILDs may require lung transplantation, a costly procedure with substantial morbidity and mortality, even though two drugs (pirfenidone and nindetanib) have been shown to decrease the rate of disease progression in UIP (usual interstitial pneumonia). Stable ILDs, on the other hand, can be managed conservatively, and ILDs without overt fibrosis may be reversible with anti-inflammatory therapy [3 - 11]. Pulmonary function tests (PFTs) are used clinically to diagnose the severity of functional impairment due to ILD by demonstrating a restrictive physiology (decreased FEV1 and FVC) with impaired gas exchange (low DLCO). PFTs nonetheless are limited as these provide only a global assessment of lung physiology and cannot discern distinct ILD patterns, with widely different prognoses.

High-resolution computed tomography (HRCT) is increasingly relied upon for more accurate characterization of ILDs. It has been shown that an UIP (usual interstitial pneumonia) pattern is associated with worse prognosis than non-UIP patterns of ILD, which include NSIP (non-specific interstitial pneumonia) and CHP (chronic hypersensitivity pneumonitis) [3, 4, 5, 10]. ILD diagnosis remains challenging and requires considerable expertise, to such an extent that the ground-truth is not derived from radiology or pathology results alone, but rather from multidisciplinary discussion (MDD) consensus [3 -11]. The current classification system is prone to substantial interobserver variability and limitations, with up to 25% of patients labeled as "unclassifiable" even after thorough review by an MDD expert team. For clinical drug trials, it would be highly desirable to have robust quantitative imaging biomarkers to reliably classify ILD and subsequently objectively assess therapy response and disease evolution. Although many studies have demonstrated the diagnostic value of pattern analysis of HRCT scans [12-24], the existing methods are limited due to several reasons: 1) most investigations have been limited to subjective, non-quantitative assessment of HRCT disease patterns; 2) many quantitative investigations have focused on simple imaging features, lacking a comprehensive characterization of complex or overlapping ILD texture patterns; and 3) automated classification results have been compared only with radiologist expert diagnosis, missing the opportunity to directly correlate quantitative imaging features with patient centered outcomes such as overall survival.

Quantitative CT imaging (QCT) with whole lung segmentation, image feature selection and texture quantification has been suggested to improve disease classification [12-19], [20 - 24]. Prior work has shown that imaging derived CT metrics can outperform PFT parameters to correctly distinguish COPD from ILD patients, utilizing computer-assisted classification via machine learning techniques such as support vector machines, with 98.1% accuracy [20]. Additional publications have suggested that automated quantification of several basic textures on thoracic HRCT can discriminate between UIP and NSIP (with accuracy of 82%) [14], as well as assess temporal changes in HRCT of patients with fibrotic interstitial pneumonias [13].

Our hypotheses are the following: 1) HRCT scans in ILD patients contain enough latent structural and functional information, which QCT driven texture based algorithms can compute in order to outperform simpler methods relying solely on histogram signatures, for disease classification; 2) Clusters of textures (phenotypes) derived from HRCT may allow for better prognostication, when compared to expert consensus diagnosis (UIP vs non-UIP patterns).

## Materials and Methods

<u>Patient Selection</u>

This retrospective study design was approved by the local institutional review boards (IRB) (#821679), and a HIPAA waiver of informed consent was granted by all participants. All methods were carried out in accordance with relevant guidelines and regulations approved by IRB. Through radiology and medical record searches, we retrospectively identified 40 patients using the following inclusion criteria: a) thoracic HRCT performed with ≤1.5 mm thickness contiguous axial slices and high spatial resolution algorithms; b) presence of fibrotic interstitial lung disease patterns on HRCT (UIP – usual interstitial pneumonia, NSIP – non-specific interstitial pneumonia or CHP – chronic hypersensitivity pneumonitis) by pathology or by a combination of clinical and radiological findings (Figure 1). HRCT datasets were anonymized and transferred to an imaging-processing computer cluster. Concurrently, PFT and clinical data were obtained from the electronic medical record, anonymized, and associated with specific imaging datasets. The dataset was balanced by design: we randomly selected 20 patients with UIP and randomly selected 20 patients with NSIP or CHP patterns of ILD, for maximizing statistical power, even with a relatively small sample size.

<u>Expert Diagnosis</u>

Subjective HRCT image analysis was performed in consensus by two thoracic radiologists (two expert radiologists with 11 and 7 years of subspecialty expertise), to assess the presence, spatial distribution and severity of the following basic imaging patterns: reticulation, traction bronchiectasis, honeycombing, ground-glass opacities, consolidations, emphysema and normal parenchyma. Additional findings such as masses or pleural effusions were noted. Accordingly, each patient was classified in one of two groups: definite or probable UIP pattern, versus most compatible with non-UIP, which comprised chronic hypersensitivity pneumonitis and non-specific interstitial pneumonia patterns of ILD. Moreover, the disease severity was stratified using a semi-quantitative score of mild (less than 25% of parenchyma involved), moderate ($\geq$ 25% but $\leq$ 50% of parenchyma involved) and severe ($\geq$ 50% of parenchyma involved). The dataset was balanced, with 20 patients in the UIP pattern, and 20 patients in the non-UIP pattern. The demographic and clinical characteristics of these 40 patients are summarized in Table 1.

QCT analysis via histogram-based (HM) method

QCT analysis was initially performed using the IMBIO Lung Texture Analysis (LTA) (CALIPER) software, which is currently investigational in the United States and not FDA approved for clinical utilization. The first step of LTA is the segmentation of the lungs, followed by segmentation of the airway tree and pulmonary vessels. The next step is to apply the CALIPER [22] algorithm to the lung parenchyma, which uses computer vision-based image analysis of volumetric histogram features, and 3D morphology to classify groups of voxels (each containing 15 1x1x1 mm voxels). The detection and quantification of lung parenchymal findings is based on histogram signature mapping techniques trained through expert radiologist consensus assessment of pathologically confirmed training sets obtained through the Lung Tissue Research Consortium (LTRC). The number of voxels in the lung parenchyma classified as each of the fundamental texture categories are calculated and converted to percentages of the combined left and right lung volume, the individual lungs, and the upper, middle, and lower sextants of each of the lungs. Voxels that are identified as vessels are not included in the calculation of the lung volume. LTA then generates as output a new series of DICOM images with multi-color, semi-transparent overlays to indicate the texture categories; and a PDF report that includes a graphic summary of the quantitative results, indicating percentage of basic parenchymal image patterns (normal, ground-glass, reticulation, honeycombing, and hyperlucent), lung volumes and spatial distribution by lung zone.

<u>QCT Analysis via Texture based (TM) method</u>

Our segmentation method is a 3-dimensional, intensity-based algorithm using *K-means* clustering to properly determine cluster centers of air / lung tissue versus soft tissue attenuation, the latter which we removed from the segmented volume. CT attenuation of the lung parenchyma (measured by Hounsfield units (HU)) for a normal subject lies between -900 to -700; and interstitial fibrosis can raise the parenchymal attenuation to as high as -100 HU, making it impossible to segment the lungs solely based on intensity thresholding. Our method provides an efficient automated lung segmentation (Figure 2), which was further refined with minor (less than 10% volume changes) manual corrections by an expert radiologist on all cases where segmentation results were suboptimal (approximately 25% of the dataset). 75% of the dataset did not require any manual corrections. All segmentation results were reviewed by an expert cardiothoracic radiologist for quality control, and following minor manual corrections as above, the entire dataset was deemed optimally segmented.

QCT feature extraction was subsequently performed via an in-house lattice-based texture estimation software pipeline (Figure 3). This method, capable of capturing the texture heterogeneity of the ILD pattern [25], is based on a regular grid virtually overlaid on each CT image. Texture features are computed from the intersection (i.e., lattice) points of the grid lines within the lung, using a local cube (window) centered at each lattice point (**Appendix** provides mathematical details). Using this novel strategy, a comprehensive set of 26 imaging features from three major statistical groups of features, gray-level histogram, co-occurrence, and run-length, were computed and saved as 3D feature maps. Features were calculated using a range of window size (ROI) traversing the image from 4 mm to 20 mm. Different ROI sizes can help to assess texture information at different spatial scales, from the finest to coarsest textures, which may capture different levels of histologic changes (from fine to coarse fibrosis). These features were averaged across each ROI for representing distinct texture phenotypes.

With the 26 3D feature maps, a *K-means* clustering approach was applied to each of the 40 patients to group the voxels within the lung that share similar feature patterns (ILD sub-types). The choice of k for the number of clusters is important. We applied K=5 for the study, where 5 tissue types were modeled: normal, ground-glass, reticular, honeycombing, and hyperlucent. The clustering results for two sample window sizes 4 mm and 12 mm are shown in Figure 4. The volume ratio of each cluster (the number of pixels for each tissue cluster divided by total number of pixels) were calculated and then fed into a Support Vector Machine (SVM) model to assess QCT ability to predict UIP vs non-UIP diagnosis. Two set of covariates were used in the model; 1: solely imaging features 2: imaging features combined with relevant clinical measures such as age, gender, and severity of the disease. Then, the TM based classification models were compared to HM method for different window sizes (Figure 5).

Although the sample size is relatively small, imaging features extracted were used to build a deep learning classification model with 5-fold cross validation. A neural network based on PyTorch framework with two hidden layers and 66 nodes was used to build the network (Appendix provides details).

Survival Analysis

The retrospective nature of our dataset allowed over 7 years of follow up, enabling correlation with relevant patient outcomes such as death, development of respiratory failure requiring ICU admission or need for lung transplantation.

Using 26 extracted features for ILD patients, Cox proportional regression hazard model [26] was performed for time-to-event outcome analysis for survival assessment prediction. The C-statistic was used as a measure of predictive performance of features. Two sets of covariates were considered; 1: UIP and non-UIP pattern labels by radiologist experts; 2: QCT imaging TM features with relevant clinical features such as age, gender, and severity for each patient as additional covariates (Figure 6). To reduce the number of covariates and the potential for overfitting, the C-statistic for each feature was evaluated based on a univariable cox regression model with cross validation.

**Results**

Imaging features derived from TM and HM were fed into a SVM model to assess their performance to classify UIP vs non-UIP patients. AUCs of the SVM models for different window sizes with and without clinical measures for TM are shown in Table 2. While HM resulted in maximum accuracy of 0.64 with CVs, TM with CVs produced a maximum AUC of 0.75 for window size = 8 mm. Window size = 8 mm together with clinical measures yielded the highest AUC (0.745) (Table 2, Figure 5). The neural network model yielded 75 accuracy with 5-folds cross validation.

Furthermore, survival analysis was performed to predict prognosis, comparing a model based on UIP vs non-UIP pattern labels by expert readers; versus a model based on TM imaging features adjusting for clinical variables. The Kaplan-Meier curves based on TM classification model (window size = 8mm) as well as labels by experts are shown in Figure 6. The QCT imaging features with clinical covariates survival model resulted in statistically significant $P$-value < 0.03 along with higher C-statistics (0.73) compared to the model solely based on UIP vs non-UIP expert labels ($P$-value = 0.59, not statistically significant).

## Discussion

The rationale for applying our TM algorithms is that simple HM algorithms that were developed to identify basic CT imaging patterns of ILD may fail to identify complex and overlapping patterns of interstitial fibrosis and inflammation. It is postulated that higher order statistics derived from TM models, which specifically and more fully characterize complex imaging textures, may allow further information to be extracted from HRCT data in patients with fibrotic ILD. Our results support that hypothesis, demonstrating the statistically significant superiority of TM algorithms for classification of ILD. Furthermore, we demonstrated the value of clinical information to improve the accuracy of any classification model based on QCT metrics, as both HM and TM models improved their classification performance when adjusting for clinical variables. It is also important to emphasize that there is an optimal window (ROI) size, which in our analysis was 8 mm. Our interpretation is that, on the one hand, smaller window sizes (e.g. < 6 mm) are more susceptible to noise due to different imaging reconstruction parameters and CT scanning technique and may not capture spatial features that can only be detected over larger areas. On the other hand, larger window sizes (e.g. > 12 mm) may cause averaging of fine interstitial abnormalities and fail to capture complex texture patterns. Therefore, there is likely an optimal window size threshold, and we found out that it is approximately 8 mm.

Our second hypothesis relied on analyzing patient outcome data (time to death, time to lung transplantation, time to ICU admission with respiratory failure) via Cox survival analysis to assess how well our TM method combining both lower level and higher level texture features (such as honeycombing, reticulation, hyperlucent, consolidation, ground-glass, normal) predicts outcomes when compared to expert consensus diagnosis (UIP versus non-UIP). The TM approach could separate two distinct survival curves ($P$-value < 0.05) while expert labeling (UIP vs non-UIP patterns) resulted in no statistically significant difference. This is of utmost clinical relevance, as it not only showcases the limitations of the current ILD classification scheme, but also offers a new and more nuanced paradigm to assess the extent of structural changes in the lungs that can be captured by QCT, but is not formally incorporated into existing classification schemes. Improved prognostication may allow more aggressive management of patients that are deemed more likely to clinically deteriorate, at the same time that it may obviate potentially risky procedures such as surgical biopsies in patients that are more likely to remain stable or even improve over time.

Our study has several limitations. First, it included a relatively small sample size, due to resource constraints. However, results showed that even with the relatively small sample size, radiomics can be useful for ILD prognosis and diagnosis given the complex nature of the disease. Notwithstanding our small sample size, it is crucial to emphasize that our results are statistically significant, demonstrating that the effects we have found are large enough to pass statistical significance tests even in this small sample size, and therefore deserve further investigation. In the future, we plan to extend our analysis to a much larger cohort of ILD patients. Second, automated segmentation results required minor manual expert correction in circa 25% of patients, especially in advanced ILD with noisy CT images, supporting the need for further technical development. This can be improved with implementing new deep learning-based segmentation approaches. Nonetheless, human revision and correction can help us to overcome this technical limitation in our study. Third, not all patients had pathology confirmation of diagnosis (though this is the usual clinical practice, as biopsies are not performed in high risk patients or patients with definite UIP pattern on HRCT). Last, the retrospective study design introduces variability in imaging acquisition protocols and clinical management (though this suggests that our method is robust enough to be able to measure a signal amidst the statistical noise from real world, non-uniform HRCT datasets, in standard clinical practice).

In summary, our results suggest that QCT lung parenchymal texture biomarkers derived from volumetric HRCT data may improve diagnosis, particularly by non-thoracic radiology experts, and even more importantly, may allow better prognostication of patients with ILD, ultimately contributing to better patient care, personalized management and treatment planning, and possibly improved long-term outcomes.

**Acknowledgements:**

The authors would like to acknowledge the contributions of the following collaborators:

Lauren Pantalone, BS, for collecting clinical information

Maya A. Galperin, MD, for analyzing HRCT scans with the lead author

**Funding:**

This research was supported by a Radiological Society of North America (RSNA) Research Seed Grant to the lead author (EB), # RSD1715. RSNA approved the study design as described in the grant proposal, however, it did not have any role in data collection, data analysis and manuscript writing.

**Disclosures:**

Dr. Eduardo J. Mortani Barbosa Jr. and Dr. Warren B. Gefter have research grants from Siemens Healthineers, which are unrelated to this study.

There is no conflict of interest for these authors: Babak Haghighi, Lauren Pantalone, Despina Kontos

**Table 1**: Patient demographics, clinical characteristics, and outcome data

| | non-UIP pattern | UIP pattern |
|---|---|---|
| *Demography and Clinical characteristics* | | |
| | N = 20 | N = 20 |
| Age (years) | 60.0 (11.2) | 60.9 (8.4) |
| Gender (Female/Male %) | 45.8/54.2 | 37.5/62.5 |
| Disease Severity (Mild, Moderate, Severe %) | 33.3/50/16.7 | 37.5/37.5/25 |
| Emphysema (no/yes %) | 83.3/16.7 | 68.8/31.2 |
| *Outcome data* | | |
| Censored/ Deceased (%) | 66.7/33.3 | 37.5/62.5 |
| Biopsy (no/yes %) | 54.2/45.8 | 62.5/37.5 |

Values expressed as mean (SD) or number (%).

**Table 2:** AUCs for classification using TM with/without clinical data

| | W = 4 mm | W = 6 mm | W = 8 mm | W = 10 mm | W = 12 mm | W = 14 mm | W = 16 mm | W = 18 mm | W = 20 mm |
|---|---|---|---|---|---|---|---|---|---|
| *AUC without clinical measures* | | | | | | | | | |
| | 0.669 | 0.698 | 0.693 | 0.656 | 0.656 | 0.591 | 0.536 | 0.568 | 0.594 |
| *AUC with clinical measures* | | | | | | | | | |
| | 0.695 | 0.732 | 0.745 | 0.682 | 0.646 | 0.677 | 0.544 | 0.568 | 0.594 |

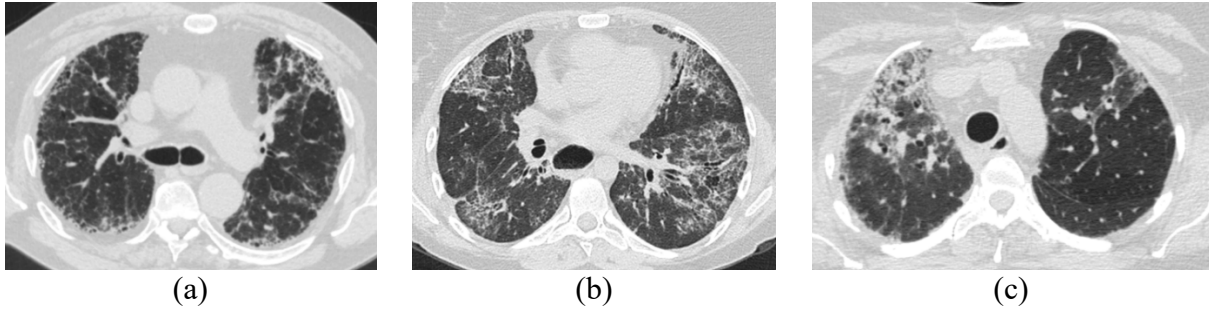

| (a) | (b) | (c) |

**Figure. 1: a)** Classic UIP (usual interstitial pneumonia) pattern, depicting symmetric, basilar and peripheral predominant textures such as reticulation, traction bronchiectasis and honeycombing. **b)** NSIP (nonspecific interstitial pneumonia) pattern, distinguished from UIP by the absence of honeycombing and the more central (peribronchovascular) distribution of reticulation and traction bronchiectasis is also noted. **c)** CHP (chronic hypersensitivity pneumonitis); the texture pattern differs from UIP and NSIP by the upper lung predominance, patchy (or non-homogeneous) distribution and presence of air trapping (noting mosaic attenuation).

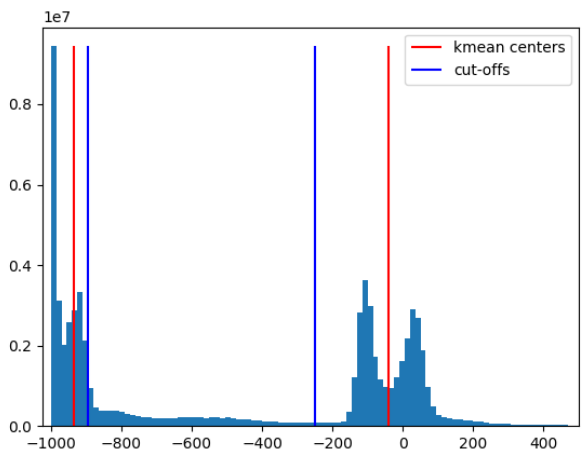

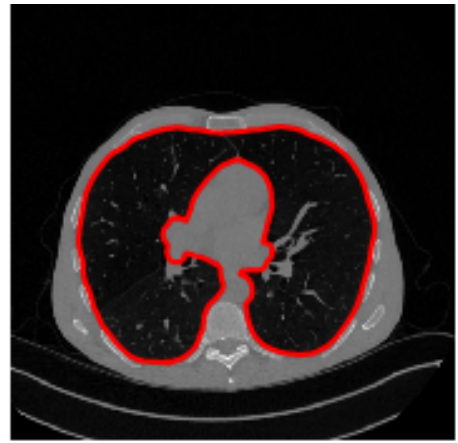

**(a)**                                               **(b)**

**Figure. 2: a)** Histogram of 3D CT image of a selected patient. Red vertical lines indicate *K-means* cluster centers identified, while blue lines are the root computed that would separate the two clusters from the object in the middle of the histogram. **b)** Resulting whole lung segmentation.

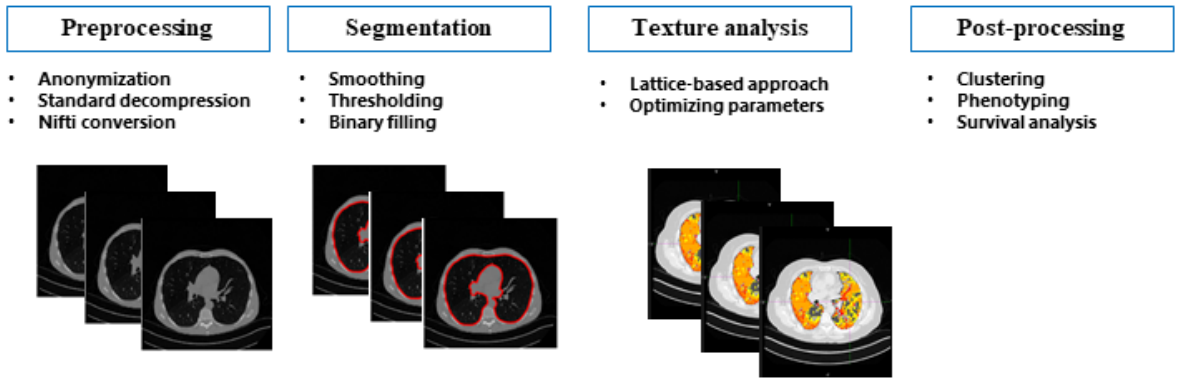

**Figure 3:** Schematic of the QCT processing pipeline, which contains pre-processing, 3D segmentation, texture analysis and post-processing steps.

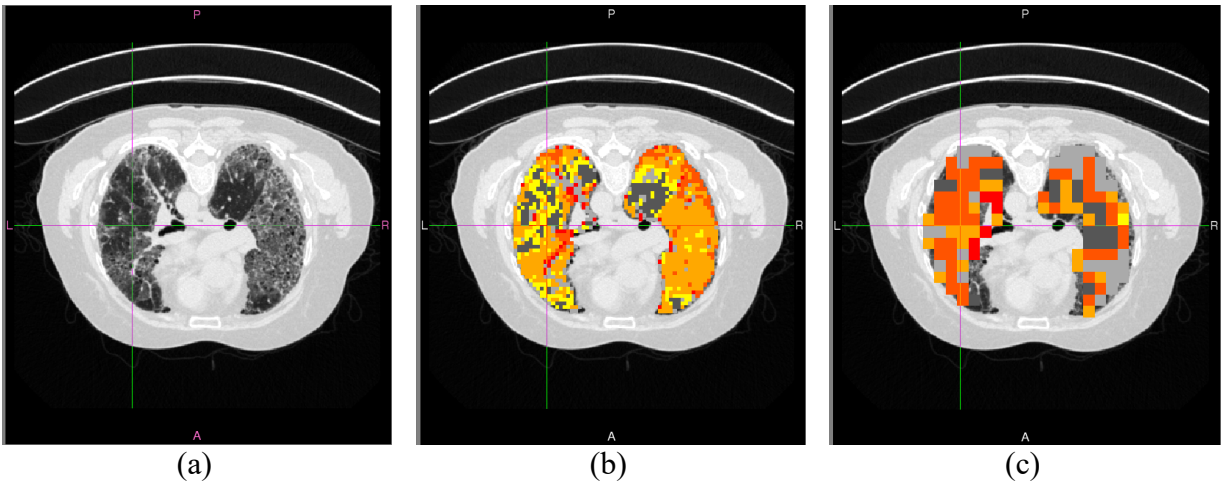

| (a) | (b) | (c) |

**Figure. 4:** Clusters resulted from 3D feature map using *K-means* clustering with different window sizes. **(a)** raw CT lung image, **(b)** window size=4mm, **(c)** window size=12mm. Five ILD subtypes (colored) including normal, ground-glass, reticular, honeycombing, and hyperlucent are shown.

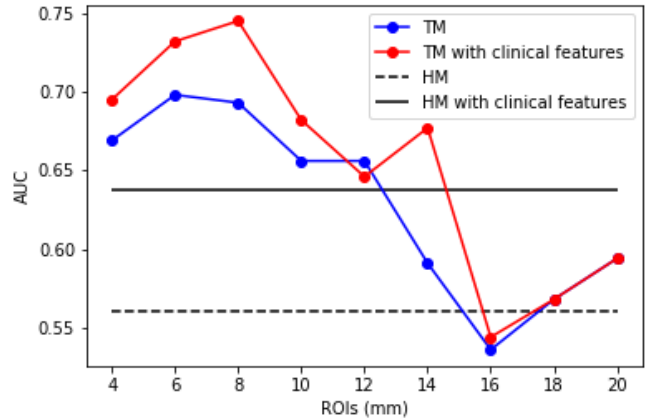

**Figure. 5:** Classification AUCs for UIP vs non-UIP labels; based on TM (blue and red curves) vs HM (dashed black and continuous black curves) features with and without clinical features including age, gender, severity, for different window sizes = 4, 6, 8, 10, 12, 14, 16, 18, 20 mm. Window size = 8 mm resulted in the best classification performance.

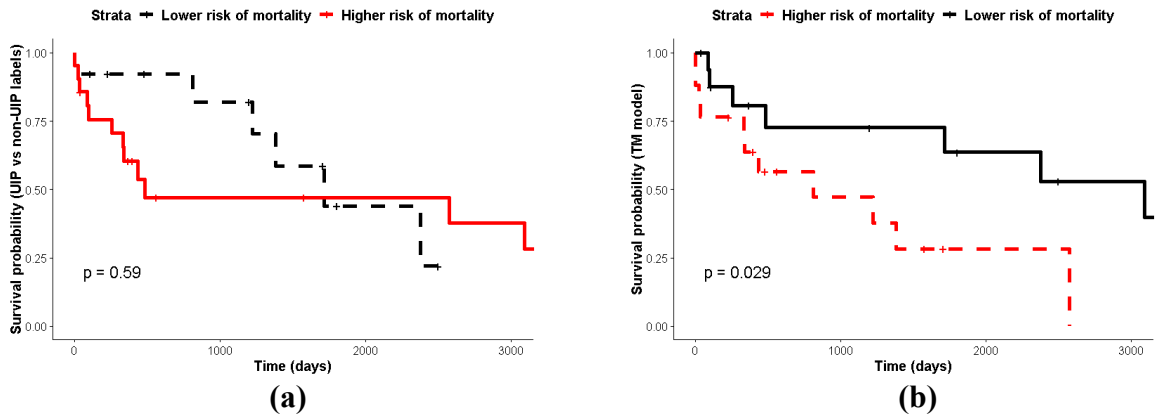

**Figure. 6: a)** Survival analysis comparing UIP vs non-UIP diagnostic labels, decided by expert consensus, demonstrate a difference in survival that is not statistically significant **b)** Survival analysis using the TM model (window size = 8mm) including relevant clinical covariates, resulting in partition of the cohort in 2 subgroups with statistically significantly different survival. *P*-values are shown for both analyses.

**APPENDIX**

Lattice-based texture estimation adopted for lung 3D CT scan was used for feature extraction. Three sets of features were derived: Gray-level histogram, Co-occurrence, Run-length features [1].

**Gray-level histogram features** are first-order histogram statistics calculated from the gray-level intensity histogram of each image.

| Feature | Mathematical Notation | Qualitative description |
|---|---|---|
| Mean | $\dfrac{\sum_k k * g(k)}{K}$ | Mean gray-level value |
| Min | $\min(k)$ | Minimum gray-level value |
| Max | $\max(k)$ | Maximum gray-level value |
| 5th Percentile | $k$: 5% of values $\leq k$ | The histogram bin that 5% of gray-level values are less than or equal to. |
| 5th Mean | $\dfrac{\sum_k k*g(k)}{\sum_k g(k)}$ for $k \leq$ 5th Percentile | Mean value of the gray-level values which less than or equal to the 5th Percentile. |
| 95th Percentile | $k$: 95% of values $\geq k$ | The histogram bin that 95% of gray-level values are less than or equal to. |
| 95th Mean | $\dfrac{\sum_k k*g(k)}{\sum_k g(k)}$ for $k \geq$ 95th Percentile | Mean value of the gray-level values which larger than or equal to the 95th Percentile. |
| Sum | $\sum_k k * g(k)$ | Sum of gray-level values |
| Sigma | $\sqrt{\sum_k (k - Mean)^2 * g(k)}$ | Variation of gray-level values around the Mean |
| Entropy | $-\sum_k g(k) * \log(g(k))$ | Measure of histogram uniformity |
| Kurtosis | $Sigma^{-4} \sum_k (k - Mean)^4 * g(k) - 3$ | Measure of histogram flatness |
| Skewness | $Sigma^{-3} \sum_k (k - Mean)^3 * g(k)$ | Measure of histogram symmetry |

where $k$ is the histogram bin, $g$ is the frequency of the histogram bin and $K = \sum_k g(k)$.

**Co-occurrence features** capture the spatial relationship between pixels and are based on the gray-level co-occurrence matrix (GLCM). In a GLCM, each element, $f(i, j)$, corresponds to the frequency with which two neighboring pixels, one with gray level $i$ and the other with $j$, occur within a specified distance ($d$). In our study, the GLCM matrices were estimated using different $window\ sizes = 4, 6, 8, 10, 12, 14, 16, 18, 20$ pixels.

| Feature | Mathematical Notation | Qualitative description |
|---|---|---|
| Cluster Shade | $\sum_{ij} \left( i - \mu_i + j - \mu_j \right)^3 * f(i,j)$ | Asymmetry in gray-level values |
| Correlation | $\sum_{ij} \dfrac{(i - \mu_i) * \left( j - \mu_j \right) * f(i,j)}{\sigma_i * \sigma_j}$ | Linear gray-level dependence |
| Haralick Correlation | $\sum_{ij} \dfrac{ij * f(i,j) - \mu_i * \mu_j}{\sigma_i * \sigma_j}$ | |
| Energy | $\sum_{ij} f(i,j)^2$ | Certainty of gray-level co-occurrence |
| Entropy | $-\sum_{ij} f(i,j) * \log \left( f(i,j) \right)$ | Uncertainty of gray-level co-occurrence |
| Inertia | $\sum_{ij} (i - j)^2 * f(i,j)$ | Local variation of gray-level intensity |
| Inverse Difference Moment | $\sum_{ij} \dfrac{f(i,j)}{1 + (i - j)^2}$ | Local homogeneity in gray-level values |

where $\mu_i = \sum_j i * f(i,j)$, $\mu_j = \sum_i j * f(i,j)$, $\sigma_i^2 = \sum_j \left( i - \mu_i \right)^2 * f(i,j)$, and $\sigma_j^2 = \sum_i \left( j - \mu_j \right)^2 * f(i,j)$.

**Run-length features** capture the coarseness of a texture in specified directions, where a run is defined as a string of consecutive pixels with similar gray-level intensity along specific linear orientation. Similar to GLCMs, a run-length matrix $R$ is defined, with each element, $R(i,j)$, representing the number of runs with pixels of gray-level intensity equal to $i$ and length of run equal to $j$ along the specific orientation. The size of the matrix $R$ is $M \times N$, where $N$ is equal to the maximum run length. The following run-length statistics were estimated using different $window\ sizes = 4, 6, 8, 10, 12, 14, 16, 18, 20$ pixels.

| Feature | Mathematical Notation | Qualitative description |
|---|---|---|
| Gray Level Non-uniformity | $\frac{1}{n_r}\sum_{i=1}^{M}\left(\sum_{j=1}^{N}R(i,j)\right)^2$ | Dissimilarity in runs across gray-level values |
| High Gray Level Run Emphasis | $\frac{1}{n_r}\sum_{i=1}^{M}\sum_{j=1}^{N}R(i,j)*i^2$ | Runs of high-gray-level values |
| Long Run Emphasis | $\frac{1}{n_r}\sum_{i=1}^{M}\sum_{j=1}^{N}R(i,j)*j^2$ | Emphasis on the long runs |
| Low Gray Level Run Emphasis | $\frac{1}{n_r}\sum_{i=1}^{M}\sum_{j=1}^{N}\frac{R(i,j)}{i^2}$ | Runs of low-gray-level values |
| Run Length Non-uniformity | $\frac{1}{n_r}\sum_{j=1}^{N}\left(\sum_{i=1}^{M}R(i,j)\right)^2$ | Dissimilarity in runs across lengths |
| Run Percentage | $\frac{n_r}{\#pixels}$ | Homogeneity and distribution of runs |
| Short Run Emphasis | $\frac{1}{n_r}\sum_{i=1}^{M}\sum_{j=1}^{N}\frac{R(i,j)}{j^2}$ | Emphasis on the short runs |

where $n_r = \sum_{i=1}^{M}\sum_{j=1}^{N}R(i,j)$ is the total number of runs.

**Deep Learning Classification**

The multi-layer neural network consists of two hidden layers with 64 nodes was used for UIP vs non-UIP classification. The input imaging features was normalized and was then processed layer by layer from the first to the final output. The results showed 75% accuracy with 5-folds cross validation for window size=4mm.

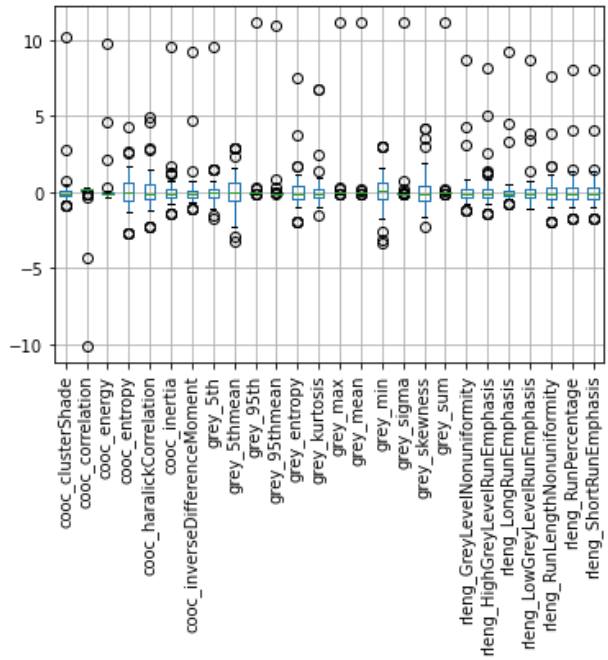

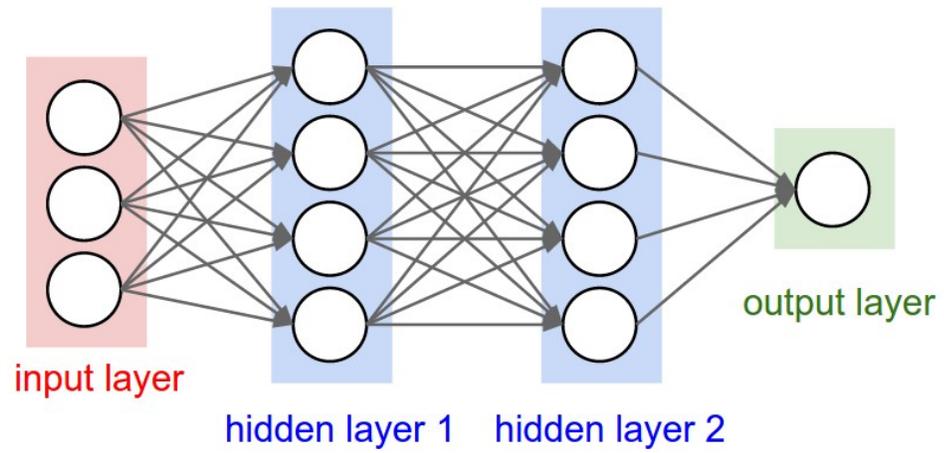

Schematic of the neural network used for classification with input layer (imaging features) and output (UIP vs non-UIP).